# Loss of charge in a Lienard-Wiechert potential implies a hidden contraction source.

J.F. Geurdes, C. vd Lijnstraat 164, 2593NN, Den Haag, han.geurdes@gmail.com

*Abstract:*

In the paper it is demonstrated that a point-like dust in a Lienard-Wiechert potential under the constraint of axial gauging must change its velocity 'beyond' a mere change in direction when some of its charge is lost at a point in space-time. The existence of a postulated hidden entity that introduces a Lorentz contraction, $\gamma'' = 1/\sqrt{1-\beta''^2}$, avoids the contradiction that hides in repeated experimentation with charge-loss in point-like dust. It is assumed that the change in velocity is 'jump-like' such that no acceleration or deceleration occurs.

*Introduction:*

Let us start with a Lagrangian density form that can also be found in Kallen's Quantum Electrodynamics (G. Kallen, 1972). We have

$$\mathcal{L} = -\frac{1}{4} F_{\mu\nu} F_{\mu\nu} - \frac{1}{2}\left(\frac{\partial A_\mu}{\partial x_\mu}\right)\left(\frac{\partial A_\nu}{\partial x_\nu}\right) = \mathcal{L}_0 + \mathcal{L}_1 \qquad (1)$$

We will use the convention here that repeated indices imply summation over, {0,1,2,3}. It is also assumed that the metric tensor has the following signature (1,-1,-1,-1) and $x_0 = ct$, while, $\vec{x}$, is the spatial three vector.

The density in (1) is an adaptation from the usual one that leads to the Maxwell equations. Suppose, a gauge different from the Lorenz-Lorentz gauge condition is selected, i.e. the second term, $\mathcal{L}_1$, is kept. We then may see from the Euler-Lagrange equation,

$$\frac{\partial}{\partial x_\nu}\left(\frac{\partial \mathcal{L}}{\partial \left(\frac{\partial A_\mu}{\partial x_\nu}\right)}\right) = \frac{\partial \mathcal{L}}{\partial A_\mu} \qquad (2)$$

that

$$\frac{\partial}{\partial x_\nu} F_{\mu\nu} = \frac{\partial}{\partial x_\mu}\left(\frac{\partial A_\lambda}{\partial x_\lambda}\right) \qquad (3)$$

In a sense, the term on the right hand side of (3) can be seen as a charge-current density, arising from not using the Lorenz-Lorentz gauge.

The retarded Lienard-Wiechert potential was originally derived in the Lorenz-Lorentz gauge but is valid in other gauges as well. That is: *we are free to choose the gauge in whatever way we like and still obtain the same physical results* (B. Thide, 2004).

In the present study, the axial gauge is taken in which, $A_3 = 0$. Note that the author will not make use of Thide's form of the Lienard Wiechert potential because of comments on the general correctness of similar derivations made by Field (J. H. Field, 2007).

The correct Lienard-Wiechert potential can, according to Field and Whitney (C. Whitney, 1988), be written like,

$$A_\mu = \frac{Q\gamma\beta_\mu}{r} \qquad (4)$$

Here, Q, is the charge, $\gamma$, the Lorentz contraction factor, $\gamma = 1/\sqrt{1-\beta^2}$, $\beta^2 = |\vec{\beta}|^2$ and $\vec{\beta} = \vec{v}/c$. Moreover, the axial gauge allows us to inspect only $\mu \in \{0,1,2\}$, which is conceptually simpler. The entity, r, is defined as a proper length, that is a length measure in the rest-frame. The charged 'dust' is supposed to move through 2-space and time in the potential given by (4). If, R, is the distance from the source to the observer, r, from (4) can also be written (C. Whitney, 1988) like

$$r = \sqrt{R_\perp^2 + \gamma^2 R_\parallel^2} \qquad (5)$$

Subsequently, the charge-current density arising from axial gauging allows

$$f(x|Q) = \left(\frac{\partial A_\lambda}{\partial x_\lambda}\right) = Q\gamma\beta_\mu \frac{\partial}{\partial x_\mu}(1/r) \qquad (6)$$

such that, $j_\mu = \frac{\partial}{\partial x_\mu} f(x|Q)$. The addition of the charge, Q, in the notation will become clear later on. Note that in Lorentz transformation, $f(x|Q)$, is invariant under general Lorentz transformation, $\Lambda_{\mu\nu}$, while for the potential we use, $A_\nu = \Lambda_{\nu\mu} A'_\mu - \eta_{\nu\mu}\frac{\partial}{\partial x_\mu}\Phi$. Note that in the summation convention used, we get d'Alembert's equation for $\Phi$, $\left(\frac{\partial^2}{\partial x_0^2} - \nabla^2\right)\Phi = 0$. Perhaps needless to say but the Lagrangian density is invariant in this case.

*Reasons why velocity must change in loss of charge*

In the subsequent section, reasons will be given why it is not possible in the axial gauge, under Lienard-Wiechert potential conditions, to lose charge in-flight and keep the same velocity. Of course, strictly speaking, acceleration cannot be a subject in special relativity. However, if it is assumed that the change in velocity will be 'jump-like' one may remain within special relativity. Moreover, it appears to be a fairly good approximation for loss of charge in point-like dust.

In the first place, let us assume that the dust may keep its original velocity when loosing charge. The only changes then will most certainly be in, Q, and in, r. We will see, $Q' = Q - \delta Q$, and let us suppose for the moment that, $r' = r - \delta r$. Now if we may assume that, $r \gg \delta r$, the following approximation can be made

$$\frac{1}{r'} \cong \frac{1}{r}(1 + \frac{\delta r}{r}) \tag{7}$$

Also assuming that, $\delta r$, changes slowly in the coordinates, we may approximately obtain from (7)

$$\frac{\partial}{\partial x_\mu}(1/r') \cong (1 + 2\frac{\delta r}{r})\frac{\partial}{\partial x_\mu}(1/r) \tag{8}$$

This change of charge leads subsequently to the following expression

$$f(x|Q') \cong \frac{Q'}{Q}(1 + 2\frac{\delta r}{r})f(x|Q) \tag{9}$$

and because, $f(x|Q)$, is invariant in this situation, the change in charge is compensated with the change in, r. We can think of an infinite small Lorentz transformation as a consequence of charge loss and we have established that $\left(\frac{\partial A_\lambda}{\partial x_\lambda}\right)$ is invariant under Lorentz transformations. Hence, we have, $f(x|Q) = f(x|Q')$

$$\frac{Q}{Q'} \cong (1 + 2\frac{\delta r}{r}) \tag{10}$$

At first sight this seems in order because, like it was stated, the change in charge, Q, is compensated for with a change in, r. From (10) we then see that because, $Q > Q'$, the, r value decreases, such that, $\delta r > 0$.

On second thoughts, however, the 'loss of charge' experiment on point-like dust can be performed a number of times and, given excellent experimentalists, and $Q = Ne$, i.e. a multiple times the unit of charge, the following 'picture' arises.

We can perform an experiment in which a point-like dust with, $N_0$, units of charge 'decays' to, $N_1$. Subsequently, we may perform, an experiment in which appoint-like dust with, $N_1$, units of charge 'decays' to, $N_2$, while finally, an experiment in which, $N_0$, units of charge 'decay' directly to, $N_2$. Decay, meaning here 'in-flight loss' of charge. We then may write for each separate experiment with point-like dust

$$\frac{N_i}{N_j} \cong \left(1 + 2\left(\frac{\delta r}{r}\right)_{i \to j}\right) \tag{11}$$

with, i<j, in {0,1,2}. Assuming this type of experimentation is possible, the following relation then arises for the, $\delta r/r$, fractions.

$$\left(\frac{\delta r}{r}\right)_{0 \to 2} = \left(\frac{\delta r}{r}\right)_{0 \to 1} + \left(\frac{\delta r}{r}\right)_{1 \to 2}$$

provided, of course, that terms: $\left(\frac{\delta r}{r}\right)_{i \to j} \left(\frac{\delta r}{r}\right)_{k \to l}$, can be neglected to single $\left(\frac{\delta r}{r}\right)_{i \to j}$ terms.

Acknowledging that the argumentation is in approximation, this result combined with (11) leads us to a relation between , $N_i$.

$$\frac{N_0 - N_1}{N_2} = \frac{N_0 - N_1}{N_1} \tag{12}$$

From this we see that it is not possible to have, $N_0 > N_1 > N_2$, but note that this *can*, however, be implemented in any experimental sequence.

Hence, the assumption that only, Q, and, r, are affected by loss of charge cannot be maintained and we have to allow velocity changes as a consequence of loss of charge. This means that we may again start with

$$f(x|Q) = Q\gamma\beta_\mu \frac{\partial}{\partial x_\mu}(1/r)$$

and somewhere along its path through space-time, the point-like dust loses charge, such that, $Q' = Q - \delta Q$, and, in a jump without acceleration, its velocity changes too. We then have,

$$f'(x'|Q') = Q'\gamma'\beta'_\mu \frac{\partial}{\partial x'_\mu}(1/r') \tag{13}$$

Moreover, let us still take, $r' = r - \delta r$ , with, $r \gg \delta r$. The change in velocity 'find its way' to the reference frames of the point-like dust in the form of the Lorentz transformation,

$$x_\nu = \Lambda_{\nu\lambda} x'_\lambda \tag{14}$$

Now, although it has been said many times, it is good to repeat, we may write

$$\frac{\partial}{\partial x'_\mu} = \frac{\partial x_\alpha}{\partial x'_\mu} \frac{\partial}{\partial x_\alpha}$$

Hence, from (14) we see $\frac{\partial}{\partial x'_\mu} = \Lambda_{\alpha\mu} \frac{\partial}{\partial x_\alpha}$. This, in turn, leads to the rewriting of (13) as

$$f'(x'|Q') = Q'\gamma'\beta'_\mu \Lambda_{\nu\mu} \frac{\partial}{\partial x_\nu}(1/r') \tag{15}$$

Subsequently, we may inspect the possibility, $\beta'_\mu \Lambda_{\nu\mu} = \beta_\nu$. But this does not lead to a result different from the conflicting previous result. The reason is, $1 - \beta^2 = \beta_\mu \beta_\nu \eta_{\mu\nu}$, such that,

$$1 - \beta^2 = \beta'_\sigma \Lambda_{\mu\sigma} \beta'_\lambda \Lambda_{\nu\lambda} \eta_{\mu\nu} = \beta'_\sigma \beta'_\lambda \eta_{\sigma\lambda} = 1 - \beta'^2,$$

when, $\Lambda_{\mu\sigma}\eta_{\mu\nu}\Lambda_{\nu\lambda} = \eta_{\sigma\lambda}$. The LT employed here does not change the 'length' of the velocity vector, $\beta^2 = |\vec{\beta}|^2$. This entails, $\gamma = \gamma'$ and note that the spatial vector in this case only is 'rotated' in three-dimensional space as a consequence of the loss of charge in a point like dust.

Hence, we may conclude that 'more than a rotation' of the three-dimensional velocity vector must occur in loss of charge in order to avoid the contradiction hidden in (12). Having said that, let us inspect e.g. the following relation,

$$\gamma'\beta'_\mu \Lambda_{\nu\mu} = \gamma''\beta_\nu \qquad (16)$$

Here, $\gamma'' = 1/\sqrt{1-\beta''^2}$. The $\beta''^2 = |\vec{\beta''}|^2$, is a hypothetical velocity vector that is introduced here in order to avoid contradiction when point-like dust loses some of its charge while traveling through space-time. If we write, $Q'' = Q'$, this idea then leads us to

$$f'(x'|Q') = Q''\gamma''\beta_\nu \frac{\partial}{\partial x_\nu}(1/r') \qquad (17)$$

Subsequent employment of (8), i.e. $\frac{\partial}{\partial x_\mu}(1/r') \cong (1 + 2\frac{\delta r}{r})\frac{\partial}{\partial x_\mu}(1/r)$, the following expression can be found

$$f'(x'|Q') \cong \frac{Q''\gamma''}{Q\gamma}(1 + 2\frac{\delta r}{r})\{Q\gamma\beta_\nu \frac{\partial}{\partial x_\nu}(1/r)\} \qquad (18)$$

The term in wavy brackets, obviously, indicates $f(x|Q) = Q\gamma\beta_\mu \frac{\partial}{\partial x_\mu}(1/r)$. Hence, we may arrive at the following

$$f'(x'|Q') \cong \frac{Q''\gamma''}{Q\gamma}\left(1 + 2\frac{\delta r}{r}\right) f(x|Q) \qquad (19)$$

Because of relativistic invariance of the, $f(x|Q)$, which arises from the invariance of, $\left(\frac{\partial A_\lambda}{\partial x_\lambda}\right)$, it follows that

$$\frac{Q\gamma}{Q''\gamma''} \cong (1 + 2\frac{\delta r}{r}) \qquad (20)$$

and this equation indeed avoids the contradiction hidden in (12). This latter claim can be demonstrated because from (20) it follows that

$$\frac{N_i}{N_j} \cong \frac{\gamma''_j}{\gamma_i}\left(1 + 2\left(\frac{\delta r}{r}\right)_{i \to j}\right) \qquad (21)$$

with , i<j, in {0,1,2}. Note that, $\gamma'' = 1/\sqrt{1-\beta''^2}$, is incontrollable when some unknown hidden cause can be attributed. But despite that, even if the experimentalists succeed in getting, $\gamma''_1 = \gamma_1$, such that $\left(\frac{\delta r}{r}\right)_{0\to 2} = \left(\frac{\delta r}{r}\right)_{0\to 1} + \left(\frac{\delta r}{r}\right)_{1\to 2}$ reappears, then still the contradiction hidden in (12) can be avoided. For, (12) then looks like,

$$\frac{N_0\gamma_0 - N_1\gamma_1''}{N_2\gamma_2''} = \frac{N_0\gamma_0 - N_1\gamma_1''}{N_1\gamma_1''} \tag{22}$$

If we subsequently suppose that, $N_0\gamma_0 - N_1\gamma_1'' \neq 0$, then, of course, $N_0 > N_1$, is possible and, $\gamma_0 \neq \gamma_1''$. In addition, we may also have, $N_1 > N_2$, together with the necessary, $N_2\gamma_2'' = N_1\gamma_1''$ and observe that, $\gamma_2'' \neq \gamma_1''$. Similar reasoning will apply when we see, $N_0\gamma_0 - N_1\gamma_1'' = 0$. We may observe, $N_0 > N_1$, and have, $\gamma_0 \neq \gamma_1''$. Hence, with the use of the velocity transformation in (16), we can have the inequality, $N_0 > N_1 > N_2$ that can be employed in experimentation.

*Conclusion*

In this paper it was demonstrated that loss of charge for a point-like dust under 'proper' Lienard-Wiechert potential conditions in the axial gauge will necessitate a change in velocity vector. This change must go 'beyond' a mere rotation of the velocity vector. The assumption of a hidden velocity component that gives rise to a Lorentz contraction factor, $\gamma'' = 1/\sqrt{1-\beta''^2}$, avoids the contradiction. The author would like to point out that a hypothetical hidden source for a contraction factor is, of course, 'atypical' physics. However, the author has demonstrated (Geurdes, 2008, Geurdes 2007) that, even in quantum mechanics, the existence of hidden factors is not that remote from reality as is commonly claimed. If the point-like dust in the present 'theory' is replaced by, for instance, a heavy ion that behaves like a 'classical particle' that in flight suffers from a partial internal neutralization of a fraction of its total charge (e.g. beta minus decay that alone cannot substantially influence the velocity of the heavy ion itself), then no external charge carrying particle can be invoked to explain the occurrence of, $\gamma'' = 1/\sqrt{1-\beta''^2}$. Because a similarity between Maxwell's field equation and Dirac's relativistic quantum mechanical equation was demonstrated previously by the author (Geurdes, 1995), it is interesting to find out what the consequences can be for relativistic quantum mechanics.

*References:*

Kallen, G, **Quantum Electrodynamics**, Springer, Berlin, N.Y. (1972).

Whitney, C. K., **On the Lienard-Wiechert potentials**, Hadronic Journal, 11, 257, (1988).

Field, J. H., **Retarded electric and magnetic fields of a moving charge**, arXiv:07041574v2, (2007).

Thide, B., **Electromagnetic Field Theory**, Upsala, 2004.

Geurdes, J. F., **Relation between relativistic quantum mechanics and classical electromagnetic field theory**, Physical Review E, 51, 5151, (1995).

Geurdes, J.F., **On the violation of the CHSH network model inequality,** arXiv:0811.1746v2 (2008)

Geurdes, J.F., **Bell's theorem refuted with a Kolmogorovian counterexample**, Int. J Theoretical Phys., Grp. Theor. & Nonl. Optics 12(3), p215-228, 2008.